\def\beg{\begin{equation}}
\def\eeq{\end{equation}}
\documentstyle[12pt]{article}
\begin{document}
\begin{center}
{\Large{\bf Quantum Gunn effect: Zero-resistance state in 2D
 electron gas.}}
\vskip0.35cm
{\bf Keshav N. Shrivastava}
\vskip0.25cm
{\it School of Physics, University of Hyderabad,\\
Hyderabad  500046, India}
\end{center}

Usually, the conductivity is quantized as the inverse of the
 resistivity, $\rho=hc/ie^2$, $\sigma=ie^2/hc$, and the velocity
versus the electric field is linear,$v=\mu E$, where $\mu$ is the
 mobility of the electrons. However, when the applied electric 
field exceeds a certain value, microwaves are emitted and the 
relation 
 $v=\mu E$ breaks down so that the velocity actually reduces as
 $E$ increases. In this region, when magnetic field is applied, 
the conductivity quantizes like the magnetic field, i.e., in 
multiples of $hc/e$ which is different from the usual quantization.
 Because of the flux quantization, the resistivity will touch zero
 in the region of 
high electric field. Factors like 4/5 arise due to new spin
 dependence of the effective charge.
\vskip1.0cm
Corresponding author: keshav@mailaps.org\\
Fax: +91-40-2301 0145.Phone: 2301 0811.
\vskip1.0cm

\noindent {\bf 1.~ Introduction}

     Recently, Mani et al$^1$ have measured the resistivity of
 GaAs/AlGaAs which is linear at small magnetic fields and shows de
 Haas- van Alphen oscillations for fields larger than (4/5)B$_f$ 
where $eB_f/m^*c=\omega_c$
is the cyclotron frequency. When the sample is irradiated with 
microwaves of frequency $\sim$ 103 GHz, the linear resistivity 
below 
the field of (4/5)B$_f$ comes down and touches the zero value. We 
wish to understand this reduction in the resistivity and why it 
approaches zero value. Zudov et al$^2$ also found that the diagonal 
resistivity touches the zero value when sample is irradiated with
 microwaves of frequency $\sim$ 57 GHz. We have reported$^3$ that
 spin determines 
the fraction which appears before the magnetic field, B$_f$ and flux 
quantization
is important to get zero resistivity. Volkov$^4$ and Bergeret et 
al$^5$ find that Gunn effect is important to obtain negative 
resistivity which is the cause of reduction with respect to the 
value without the microwave radiation and this type of negative
 resistance is in agreement with the experimental work of
 Willett$^6$ where observation of the negative resistance was 
mentioned. 

     At small electric fields, the velocity is linearly related to
 the electric field, $v=\mu E$. However, at large fields, the 
velocity does not increase with increasing field but it decreases. 
In this region, 
it is interesting to quantize the conductivity. Since the system 
radiates microwaves, it is called the Gunn effect. The quantization 
of the magnetic field in the region of Gunn effect leads to 
quantized conductivity. Thus, we report the ``quantum Gunn effect" 
(QGE) which
 is the quantized conductivity in the region of electric field
 relevant to Gunn effect. Usually, the conductivity is quantized 
in units of $e^2/hc$
but in the present case, it is quantized in units of hc/e with a 
suitable multiplier so that the units are correct. Another aspect of 
the QGE is that as $n$ is varied, several microwave frequencies 
are emitted instead of only one frequency of the Gunn effect.

\noindent{\bf 2.~~Theory}

     Usually the conductivity, $\sigma$, is defined by, $j=\sigma E$
where $j$ is the electrical current and E is the electric field. 
The current is defined in terms of velocity,$v=\mu E$ where $\mu$ is 
called the mobility. The electron as well as the hole currents can 
thus be written in terms of velocity  as,
\beg
j=n_iev_e+n_hev_h=n_ie\mu_iE+n_he\mu_hE
\eeq
This means that $v$ is a linear function of E. Actually, at large
 electric fields, $v$ as a function of $E$ ceases to be linear and 
when there is sufficient heating, energy is emitted in the form of
 microwaves. Such an emission of microwaves for electric fields 
larger than a certain characteristic value, $E_o$ is called the 
Gunn effect. For $E>E_o$, the velocity actually decreases with 
increasing $E$. Therefore, the region, $E>E_o$, offers an unusual 
opportunity to investigate the flux quantization. The usual 
quantization of 
resistivity or that of conductivity, is based on the region $E<E_o$,
 where $\rho=h/ie^2$ or $\sigma=ie^2/h$. We therefore expect a new 
type
 of quantization for fields $E>E_o$. The number of particles with 
velocity in the range of $dv$ at $v$ is given by Maxwell velocity 
distribution but conversion of velocity into electric field requires 
the knowledge of linearity between the velocity and the electric 
field. For $E>E_o$, such a linear formula is not valid. Therefore,
 we use a simple  form which has a peak at $E_o$. For $E<E_o$, this 
function has a linear region so that velocity increases with 
increasing electric field upto $E=E_o$. Once $E>E_o$, the linear 
behavior is not needed and the system emits microwaves. Therefore,
 the distribution should have a peak at $E_o$ and should fall for
 $E>E_o$. Therefore we use,
\beg
v=\mu E {\Delta^2\over (E-E_o)^2+\Delta^2}
\eeq
instead of $v=\mu E$. At $E=E_o$, there is a peak in the velocity. 
For $E<E_o$, the velocity becomes,
\beg
v=\mu E{\Delta^2\over E_o^2+\Delta^2}
\eeq
which is linear in $E$ but is smaller than $\mu E$. At the peak 
$E=E_o$, $v=\mu E$ is established and for $E>E_o$,
$v=\mu E[\Delta^2/(E^2+\Delta^2)]$ which falls with increasing $E$. 
This is 
the region where Gunn effect is known to occur. When the distribution
 of velocity as a function of electric field (2) is used the current
 becomes,
\beg
j=n_ie\mu_iE{\Delta^2\over (E-E_o)^2+\Delta^2}+n_he\mu_hE
{\Delta^2\over (E-E_o)^2+\Delta^2}
\eeq
where $n_i$($n_h$) are the electron (hole) concentrations and 
$\mu_i$ and $\mu_h$ are the electron and hole mobilities, respectively.
 The
conductivity is determined by two terms, one due to electrons,
 $\sigma_e$ and the other due to holes, $\sigma_h$, with,
\beg
\sigma_e=n_ie\mu_i\Delta^2[{1\over (E-E_o)^2+\Delta^2}-{2E(E-E_o)
\over \{(E-E_o)^2+\Delta^2\}^2}]+e\mu_i\Delta^2{dn_i\over dE}{E\over
 (E-E_o)^2+\Delta^2}
\eeq
and
\beg
\sigma_h=n_he\mu_h\Delta^2[{1\over (E-E_o)^2+\Delta^2}-{2E(E-E_o)\over 
\{(E-E_o)^2+\Delta^2\}^2}]+\mu_he\Delta^2{dn_h\over dE}{E\over
 (E-E_o)^2+\Delta^2}
\eeq
When the electric field is increased, if the electron concentration
 increases, that of the holes must reduce so that,
\beg
{dn_e\over dE}= - {dn_h\over dE}
\eeq
Substituting this result in (5) and (6) and adding the two terms,
\beg
\sigma= \sigma_e+\sigma_h=\sigma_o+(\mu_i-\mu_h)e\Delta^2{dn_i\over
 dE}{E\over(E-E_o)^2+\Delta^2}
\eeq
where
\beg
\sigma_o=e\Delta^2(n_i\mu_i+n_h\mu_h)[{1\over (E-E_o)^2+\Delta^2}-
{2E(E-E_o)\over\{(E-E_o)^2+\Delta^2\}^2}].
\eeq
In the Gunn effect, microwaves are emitted once the applied 
electric field is larger than $E_o$ but there is no need of
 any magnetic field. There are many calculations of the application
 of crossed electric and magnetic fields to the semiconductors but 
in the present problem, $v=\mu E$ has been avoided and peaked
 distribution has been used so that it will be of interest to see 
the effect of magnetic field when $E>E_o$.
We can apply the magnetic field by considering the Lorentz force so
 that the effective electric field becomes,
\beg
E'=E+{1\over c}(\vec v \times \vec B).
\eeq
In the present case $\vec v$ is a peaked function so that the electric
 field becomes,
\beg
E'=E+{1\over c}\mu(\vec E\times \vec B){\Delta^2\over (E-E_o)^2+
\Delta^2}
\eeq
where $\mu$ is the mobility.The effective electric field then becomes,
\beg
E'=|E|[1+{1\over c}{\mu \Delta^2 B\over (E-E_o)^2+\Delta^2}]
\eeq
The second negative term of (9), upon the application of magnetic 
field to electrons becomes,
\beg
\sigma_{o-}\simeq {-2e\Delta^2n_i\mu_i|E|^2\over \{(E-E_o)^2+
\Delta^2\}^2}[1+{1\over c}{\mu_i\Delta^2 B\over (E-E_o)^2+\Delta^2}]
[1-{E_o\over|E|}]
\eeq
which shows that when the applied field is larger than certain value, 
the conductivity can reduce. We can write the above conductivity as,
\beg
\sigma_{o-}\simeq -a_1[1+a_2 B]
\eeq
Where we consider the flux quantization, so that the magnetic field 
in the area A can be written as B.A=n$\phi_o$ where $\phi_o$=hc/e
 which means that the conductivity in the Gunn region becomes,
\beg
\sigma_{o-}\simeq -a_1[1+a_2n\phi_o/A]\simeq -a_1[1+a_2(n/A)hc/e]
\eeq
This formula is very different from the $ie^2/h$ type quantization of 
conductivity but this is a ``quantized Gunn effect". Without the 
quantization, where only a single frequency microwave was emitted,
 now many more microwave frequencies will be emitted due to different 
values of $n$ in (15). However, the expression (13) is quite 
approximate so that actual behaviour is much more complicated than by 
only one term 
in (15).

     Thus we learn that conductivity has positive terms and electric
 field dependent negative terms so that when the electric field is 
increased the negative term in the conductivity can become large so 
that conductivity reduces. Thereafter, upon application of magnetic
 field, plateaus can arise due to ``flux quantization" This 
conductivity quantization is different from the usual quantization 
in terms of $ie^2/h$. For different values of the integer several
 plateaus can arise which means that instead of a single microwave 
frequency emission in the Gunn effect, several frequencies can arise. 
These predicted features of the  conductivity are qualitatively 
in accord with the present day experimental data. The prediction of
 emission of multiple freuencies in the Gunn effect has not yet been
 varified by the experiments. However, such multiple frequencies in 
the Gunn effect must occur. The flux quantization measures the product
 of the charge and the magnetic field. Therefore, fractional numbers
 arise in the product $eB$. These fractions are well predicted in a 
previous paper$^3$ where it was found that the flux quantization 
depends on
 spin.

\noindent{\bf3.~~ Discussions}.

     Anderson and Brinkman$^7$ have pointed out the importance of
 observation of zero resistivity when the sample is irradiated with 
microwaves at a frequency somewhat higher than the cyclotron frequency.
They have suggested that it depends on the structure of the energy
 levels in crossed electric and magnetic fields. The old calculations
 of the crossed electric and magnetic fields have positive resistance.
 We have found that the factors like 4/5 arise only when spin is
 involved due to the  spin dependent flux quantization. The resistivity
 is zero for the flux-quantized state. The negative resistance arises 
due to heating of the sample above an electric field of $E_o$. This
 negative resistance phenomenon is called the Gunn effect where 
microwaves are emitted by the same. The emission of radiation is 
associated with negative resistance with absorption associated with 
positive resistance. The microwave frequency somewhat higher than the
 cyclotron frequncy is thought to be due to scattering by impurities
 by the Yale group$^8$. However, there is no need of impurities in our
 calculation and we obtain the correct values without impurities in 
the pure sample$^{9,10}$.

     Cheremisin$^{11}$ has correctly pointed out that $j=\sigma_{xy}E$ 
is violated. In the N shaped current(voltage) characteristics, if peak
 occurs for negative $E$, then it serves not much purpose.  Similarly, 
in the S-shaped I(V) characteristics there is no peak for positive
 electric field. We need a peak in the current(voltage) characteristics
 for positive value of the electric field as we have introduced by 
eq.(2). In this case Gunn effect leads to reduced or negative
 resistivity needed by the experiment. Similarly, Shi and Xie$^{12}$ 
used the current linear in $E_o$
along with integral of current-current correlations. This will not
 produce a Gunn effect which we obtain because of peaked relationship
 between velocity and the electric field.

     When microwaves are emitted by the Gunn effect$^{13-18}$, it is 
important that Goldstone theorem is obeyed. That means that Goldstone 
bosons are emitted. These Goldstone bosons are the heat waves or 
phonons. The emission of phonons must be accompanied by charge-density
 waves. This can be seen, for example, by factorizing the
 electron-phonon interaction. Making factors of electron number
 density automatically produces soft phonons. Therefore, there are
 charge density waves which actually emit electromagnetic waves or
 microwaves which is another way of saying that there is Gunn effect
 which leads to negative or reduced resistance. We can obtain zero
 resistance by having spin singlets as in superconductivity or 
triplet states as in $^3$He. The charge density waves are
 paramagnetic and hence have small resistance. When energy is emitted,
 we call it ``negative resistance". In the present phenomenon, 
there are spin singlets with one component leading to zero resistance 
accompanied by flux quantization.

\noindent{\bf4.~~ Conclusions}.
    
     We propose a flux-quantized Gunn effect as the interpretation
 of negative resistance which also has plateaus. The fractions 
before the cyclotron frequency arise due to ``spin-dependent flux
 quantization". The zero-resistance plateaus are due to flux 
quantization. The zero as well as negative resistance thus is well
 understood. There is a striking similarity between the resistance
 in the present samples and the diagonal resistance in the quantum
 Hall effect. In fact, in the correct theory$^9$ which produces the 
same fractional charges as the experimental data, the requirement 
of the Hall geometry is not strictly needed. The fractional charge
 depends on the spin and the orbital state of the quasiparticles.
 The fractional charges, such as 4/3, found in the data are the same
 as in the calculated tables. As far as the emission of microwaves 
is concerned, more than one frequency will be generated due to 
integer, $n$ in $n\phi_o$ which quantizes the magnetic field.
\vskip1.25cm

\noindent{\bf5.~~References}
\begin{enumerate}
\item R. G. Mani, et al, Nature {\bf419}, 646(2002)
\item M. A. Zudov, et al,  Phys. Rev. Lett. {\bf 90}, 046807(2003).
\item K. N. Shrivastava, cond-mat/0303309.
\item A. F. Volkov, cond-mat/0302615.
\item F. S. Bergeret, B. Huckestein and A. F.. Volkov, cond-mat/0303530.
\item R. L. Willett, K. W. West and L. N. Pfeiffer, Bull. Am. Phys. Soc.
 H23.001(March 2003).
\item P. W. Anderson and W. F. Brinkman, cond-mat/0302129.
\item A. C. Durest, S. Sachdev, N. Read and S. M. Girvin,
 cond-mat/0301569
\item K. N. Shrivastava, cond-mat/0212552.

\item K.N. Shrivastava, Introduction to quantum Hall effect,\\ 
      Nova Science Pub. Inc., N. Y. (2002)ISBN 1-59033-419-1.\\
\item M. V. Cheremisin, cond-mat/0304581.
\item J. Shi and X. C. Xie, cond-mat/0302393.
\item M. P. Shaw, H. L. Grubin and P. R. Solomon, The Gunn-Hilsum 
effect,
Academic Press, New York 1979.
\item J. B. Gunn, Solid State Commun. {\bf 1}, 88 (1963).
\item J. B. Gunn, IBM J. Res. Dev. {\bf 10}, 300(1966).
\item C. Hilsum,Proc. IRE {\bf 50}, 185(1962).
\item P. N. Butcher and W. Fawcett, Proc. Phys. Soc. (London)
 {\bf 86},1205(1965).
\item H. Kroemer, Proc. IEEE {\bf 52}, 1736 (1964).
\end{enumerate}
\vskip0.1cm

\end{document}